\documentclass[%
 reprint,
 superscriptaddress,
 showpacs,preprintnumbers,
 amsmath,amssymb,
 aps,
 prb,
]{revtex4-1}

\usepackage{graphicx}
\usepackage{dcolumn}
\usepackage{bm}

\usepackage{upgreek}
\usepackage{amsmath}

\begin{document}


\title{Manipulating strong coupling between exciton and quasi-bound states in the continuum resonance}

\author{ Meibao Qin }
\affiliation{School of Physics and Materials Science, Nanchang University, Nanchang 330031, People’s Republic of China}

\author{Junyi Duan}
\affiliation{Institute for Advanced Study, Nanchang University, Nanchang 330031, People’s Republic of China}
\affiliation{Jiangxi Key Laboratory for Microscale Interdisciplinary Study, Nanchang University, Nanchang 330031, People’s Republic of China}

\author{Shuyuan Xiao}
\email{syxiao@ncu.edu.cn}
\affiliation{Institute for Advanced Study, Nanchang University, Nanchang 330031, People’s Republic of China}
\affiliation{Jiangxi Key Laboratory for Microscale Interdisciplinary Study, Nanchang University, Nanchang 330031, People’s Republic of China}
\author{Wenxing Liu }
\affiliation{School of Physics and Materials Science, Nanchang University, Nanchang 330031, People’s Republic of China}
\author{Tianbao Yu }
\email{yutianbao@ncu.edu.cn}
\affiliation{School of Physics and Materials Science, Nanchang University, Nanchang 330031, People’s Republic of China}
\author{ Tongbiao Wang }
\affiliation{School of Physics and Materials Science, Nanchang University, Nanchang 330031, People’s Republic of China}
\author{ Qinghua Liao }
\affiliation{School of Physics and Materials Science, Nanchang University, Nanchang 330031, People’s Republic of China}
\begin{abstract}
Strong coupling exhibits unique ability to preserve quantum sates between light and matter, which is essential for the development of quantum information technology. To explore the physical mechanism behind this phenomenon, we employ the tight-binding method for expanding the temporal coupled-mode theory, with the absorption spectrum formula of coupled system directly obtained in an analytical way. It reveals all the physical meaning of parameters defined in our theory, and shows how to tailor lineshapes of the coupled systems. Here, we set an example to manipulate the strong coupling in a hybrid structure composed of excitons in monolayer WS$_2$ and quasi-bound states in the continuum  supported by the TiO$_2$ nanodisk metasurfaces. The simulated results show that a clear spectral splitting appeared in the absorption curve, which can be controlled by adjusting the asymmetric parameter of the nanodisk metasurfaces and well fitted through our theoretical predictions. Our work not only gives a more comprehensive understanding of such coupled systems, but also offers a promising strategy in controlling the strong light-matter coupling to meet diversified application requests.
\end{abstract}

\maketitle


\section{\label{sec1}Introduction}

Enhancing the light-matter interaction could offer a large variety of applications in many areas, such as quantum electrodynamics, nonlinear optics, photonics, and so on\cite{Zheludev2012,Xiao2020a}. Strong coupling occurs when the coherent exchange rate between the two subsystems with the coincident frequencies exceeds their individual decay rate, and results in the formation of a new hybrid light-matter state\cite{Deng2010,Torma2014,Dovzhenko2018,Huang2022,AlAni2022}. A key aspect of achieving this lies in enhancing the coupling strength $g$, which can be determined as $g = \mu  \cdot E \propto \frac{1}{V}$ in the dipole approximation\cite{Hugall2018}. Here, $\mu$ and $E$ are the transition dipole moment and the local electric field intensity, respectively. In order to fulfill this requirement, monolayer transition-metal dichalcogenides (TMDCs) emerge as a kind of promising materials due to their large transition dipole moment and other extraordinary electronic and optical properties, such as high binding energy owing to the quantum confinement of the atomic and the direct band gaps\cite{Xia2014,Wang2018}. One of the feasible ways to enhance the strong coupling is to integrate the monolayer TMDC to the confined electromagnetic field. In this case, optical nanocavities exhibit excellent light trapping, and have been proven to be an efficient way to realize strong field enhancement. The conventional plasmonic nanostructures confine light within subwavelength mode volume but suffer heavy ohmic loss in the visible region, which limits to their practical applications\cite{Chen2012,Tran2017}. In contrast, dielectric nanostructures do have low intrinsic loss, and allow the existence of the multiple interference in resonant high-index dielectric nanostructures, offering novel opportunities for the controlling of the light-matter interaction\cite{Kuznetsov2016,Koshelev2019,Koshelev2020}. In particular, a recently emerged concept of bound states in the continuum (BIC) enables a promising way to achieve strong light-matter interaction, which benefits from their better confinement of energy and robust existence\cite{Hsu2013,Koshelev2018a,Li2019}.
 
In recent years, many researchers have investigated strong coupling between the TMDC and plasmonic/dielectric nanostructures \cite{Kang2014,Zu2016,Sun2018,Zhang2018,Koshelev2018,Kang2018,Du2019,Wang2019,Li2019a,Xie2020,Cao2020,Chen2020,Xie2020a,Zhao2020,Asham2021,Qin2021}, and developed a general approach to employ the coupled harmonic oscillator model for such hybrid systems. Although this method can be exploited to deal with the interaction between the eigenmodes of the coupled system, the subject is limited to the closed cavity-exciton system, while the influence of external excitation is ignored. However, the coupled systems are open in nature, where different resonant modes couple not only with each other via near field but also with the external free space\cite{Lin2020},  therefore the existing theory cannot describe the optical response of the actual coupled system, i.e., transmission/absorption spectrum. As an early attempt, the existing temporal coupled mode theory (TCMT) can deal with the coupling between propagation modes and background modes. Unfortunately, it cannot resolve the problem that bound states coupled to various ports, such as exciton modes in strongly coupled systems. Therefore, an analytical theory that can consider the near field and external free space in the same foot is in urgent need. On this issue, the tight-binding method that can divide the wave function into propagating state and bound state, becomes an ideal platform to solve such bound states. Therefore, utilizing the tight-binding method for generalizing temporal coupled mode theory is expected to be helpful to deal with the interaction among each resonant modes and with the free space as well.
 
In the present work, we derive a formal theoretical work based on the photonic tight-binding method and TCMT, to better handle the interaction between coupled system and external free space. As a proof of demonstration, we investigate a hybrid structure composed of monolayer WS$_2$ placed on top of TiO$_2$ nanodisk metasurfaces supporting magnetic dipole quasi-BIC resonance. In order to fulfill strong coupling condition, we keep the frequencies of the two coupled modes consistent by adjusting the height and asymmetric of the nanodisk metasurfaces, and the couple strength $g=17.91$ meV can be achieved through benchmark calculations. The typical anti-crossing behavior of the Rabi splitting 29.33 meV in the absorption spectrum can be observed, which agrees well with our proposed theory. Moreover, we applied the approach to tailor the absorption spectrum in the various regimes from weak coupling to strong coupling. Our work provides a broad and deep insight into the properties of the coupled systems, which is much closer to the actual situation, and holds a great potential for controlling strong coupling interaction. 
\section{\label{sec2}Generalized temporal coupled mode theory}
We develop the theory on the basis of the tight-binding method \cite{Raman2010,Xi2011}, to provide insight into the propagating state and bound state of a coupled system consisted of exciton in monolayer TMDC and arbitrary resonator under certain external illumination. The Hamiltonian of such open system can be written as
\begin{equation}
\hat{H}=\hat{H}_\text{b}+\hat{V_\text{n}}+\hat{V_\text{e}},\label{eq1}
\end{equation}
where \(\hat{H}_\text{b}\) is the Hamiltonian of the background. \(\hat{V}_\text{n}\) and \(\hat{V}_\text{e}\) represent the potential contributed by the resonator and exciton, respectively. Then, we need to solve the following eigen equation of the whole system
\begin{eqnarray}
	\hat{H}|\Psi\rangle&=&\omega|\Psi\rangle,\label{eq2}	
\end{eqnarray}
here \(|\Psi\rangle=|\Psi_\text{in}\rangle+|\Psi_\text{rs}\rangle\) is the total wave function with $\Psi_\text{in} $ and $\Psi_\text{rs}$ describing the incident port and resonance mode, respectively. For this open system with two resonance modes, $\Psi_\text{in} $ and $\Psi_\text{rs}$ can be further written as
\begin{eqnarray}
	|\Psi_\text{rs}\rangle&=&a_\text{n}|\Psi_\text{n}\rangle+a_\text{e}|\Psi_\text{e}\rangle,\label{eq3}	\\
	|\Psi_\text{in}\rangle&=&\sum_{q}s^+_q|\Psi_{q}\rangle,\label{eq4}	
\end{eqnarray}
where $a_\text{n}$ and $a_\text{e}$ are the strength of the electric field excited by the resonator and exciton, respectively, under external illumination described by ${s^+_q}$ denoting the excitation amplitudes at different incoming ports. Due to unavoidable radiation in open systems, we should not only consider near field (NF), but also include the far field (FF) propagating to the external continuum. $|\Psi_\text{n}\rangle$ and $|\Psi_\text{e}\rangle$ can be expressed as 
\begin{eqnarray}
	|\Psi_\text{n}\rangle&=&|\Psi^\text{NF}_\text{n}\rangle+|\Psi^\text{FF}_\text{n}\rangle,\label{eq5}	\\
	|\Psi_\text{e}\rangle&=&|\Psi^\text{NF}_\text{e}\rangle+|\Psi^\text{FF}_\text{e}\rangle,\label{eq6}	
\end{eqnarray}
with $|\Psi^\text{NF}_\text{n}\rangle$ and $|\Psi^\text{FF}_\text{n}\rangle$ representing the NF part and FF of the resonator wave function, respectively. $|\Psi^\text{NF}_{e}\rangle$ and $|\Psi^{FF}_\text{e}\rangle$ describe the NF part and FF of the exciton wave function, respectively. It is noted that the exciton mode in high refractive dielectric is a kind of bound sate. hence the far filed of its wave function can be defined as $|\Psi^{FF}_\text{e}\rangle=0$.

In order to apply the tight-binding method to manipulate strong coupling, we need to construct a coupling matrix $D$ between the resonant modes and the ports. In the proposed hybrid system, only the resonator exhibits loss rate $\gamma$, and one incident port means $q=1$ in the Eq. (\ref{eq4}), so the coupling matrix can be expressed by the form $D$=
$\begin{pmatrix}
	\sqrt \gamma  &  0 \\ 
	\sqrt \gamma  &  0 
\end{pmatrix}$. The coupling among eigen modes and the loss of the system can be expressed as 
\begin{eqnarray}
\Omega&=&\Omega _1+\Omega _2=\begin{pmatrix}
	\omega _\text{n}   &   0\\
	0           &   \omega _\text{e}
\end{pmatrix}+\begin{pmatrix}
0   &   g\\
g   &   0
\end{pmatrix},\label{eq7}\\
\Gamma&=&\Gamma _{\text{rad}}+\Gamma _{\text{abs}}=\frac{D^\dagger D}{2}+\Gamma _{\text{abs}} \notag\\
      &=&\begin{pmatrix}
	\gamma   &    0\\
	0        &    0
\end{pmatrix}+\begin{pmatrix}
	0   &   0\\
	0   &   \delta
\end{pmatrix},\label{eq8}
\end{eqnarray}
where $\Omega$ is a $2\times2$ matrix determined by the interaction of the eigenmodes. $\omega_\text{n}$ and $\omega_\text{e}$ represent the resonance frequency of the resonator and exciton, respectively. $\delta$ is the absorption loss of exciton and $g$ gives the coupling strength between resonance mode and exciton. $\Gamma$  describes the loss of the hybrid system, related by the radiation and absorption loss rates.

Based on the above derivations, we can generalize the TCMT\cite{Fan2003}, to describe the light-matter interaction in the coupled system,
\begin{eqnarray}
\frac{d}{dt}\begin{pmatrix}
	a_\text{n}\\
	a_\text{e}
\end{pmatrix}&=& i \begin{pmatrix}
\omega_\text{n}+i\gamma  &  g\\
  g               &  \omega_\text{e}+i\delta
\end{pmatrix}\begin{pmatrix}
a_\text{n}\\
a_\text{e}
\end{pmatrix}\\
&+&\begin{pmatrix}
	\sqrt\gamma  &  \sqrt\gamma\\
	0            &  0
\end{pmatrix}\begin{pmatrix}
	s^+_1\\
	0
\end{pmatrix},\label{eq9} \notag\\
|s_{-}\rangle&=&C|s_{+}\rangle+D|a\rangle,\label{eq10}
\end{eqnarray}
Here, $|s_{+}\rangle$ and $|s_{-}\rangle$ are the amplitude of plane wave in the input and output, respectively. $|a\rangle$ is the resonance amplitude of the mode with a general time dependence of $\propto e^{i\omega t}$, the dynamic equations for resonance amplitudes can be written in the following form:
\begin{eqnarray}
	i|a\rangle &=&[(\hat{I}\omega-\Omega)-i\Gamma]^{-1}D^T|s_{+}\rangle,\label{eq11}\\
    \langle a|i&=&-\langle s_{+}|(D^{T})^\dagger[(\hat{I}\omega-\Omega)+i\Gamma]^{-1}.\label{eq12}
\end{eqnarray}
According to the energy conservation, we have the relation for resonance amplitude and time,
\begin{eqnarray}
\frac{{d{{\left| a \right|}^2}}}{{\operatorname{dt} }}&=&\frac{{d\left\langle {a}
		\mathrel{\left | {\vphantom {a a}}
			\right. \kern-\nulldelimiterspace}
		{a} \right\rangle }}{{\operatorname{dt} }} \notag\\
	&=& + \left( {\left\langle a \right|\partial _t^ \dagger } \right)\left| a \right\rangle  + \left\langle a \right|{\partial _t}\left| a \right\rangle \notag\\  
	&=&2\left\langle a \right|\Gamma \left| a \right\rangle,\label{eq13}
\end{eqnarray}
and the absorption of the coupled systems at the frequency can be finally obtained, 
\begin{eqnarray}
A&=\frac{\langle S_+|(D^T)^\dagger[(\hat{I}\omega-\Omega)+i\Gamma]^{-1}\Gamma_{abs}[(\hat{I}\omega-\Omega)-i\Gamma]	
D^T|S_+\rangle}{\langle S_+|S_+\rangle} \notag\\
&=\frac{2\delta \gamma{g}^2}{[(\omega-\omega_\text{n})(\omega-\omega_\text{e})-\delta \gamma-{g}^2]^2+[\delta(\omega-\omega_\text{n})+
	\gamma(\omega-\omega_\text{e})]^2}.\label{eq14}
\end{eqnarray}

In such a formula, the absorption spectrum can be directly predicted, and will exhibit an asymmetric Fano line shape with two peaks and one dip, when satisfying the condition $g^2>\frac{\gamma^2+\delta^2}{2}>\delta\gamma$. Following this, strong coupling can be realized by the control of the radiation loss and absorption loss. 

\section{\label{sec3}Structure design and numerical model}

\begin{figure}[htbp]
	\centering
	\includegraphics
	[scale=0.36]{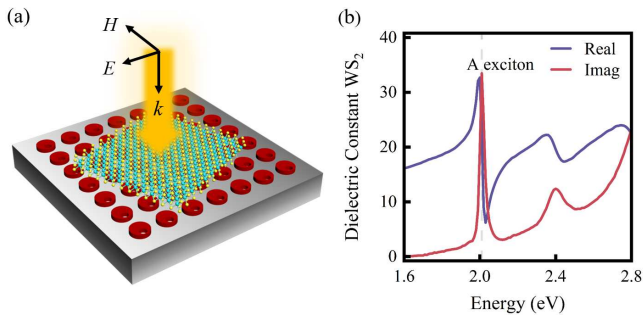}
	\caption{\label{fig1} (a) Schematic of the proposed coupled system composed of a monolayer WS$_2$ deposited on the nanodisk  metasurfaces. (b) The dielectric constant of the monolayer WS$_2$, blue line and red line represent real and imaginary parts, respectively. }
\end{figure}

In order to identify the reliability of the generalized TCMT, we consider an absorption system composed of a monolayer WS$_2$ laid flat on the array of TiO$_2$ nanodisk with the SiO$_2$ substrate, as the sketches shown in Fig. \ref{fig1}(a). Each unit cell is characterized by a square lattice with a period of
$P=410$ nm, thickness $H=85$ nm, radius is $R=140$ nm, and a nanohole with a variable radius is introduced at a fixed distance $d=70$ nm away from the center of the nanodisk. The perturbation with an asymmetric parameter $\alpha=\frac{\pi r^2}{\pi R^2}$ is thus introduced into an in-plane inverse symmetric of the structure, which can build a radiation channel and transform the symmetric-protected BIC into the quasi-BIC \cite{Xu2019,Li2019,Wang2020}.

For the material modelling, TiO$_2$ can be a good candidate owing to its high refractive index $n_{\text{TiO}_2}=2.6$ and negligible extinction coefficient in the range of the visible wavelengths, which can constitute optically-induced electric and magnetic Mie resonators and strongly confine electromagnetic energy within the structured metasurfaces\cite{Sun2017,Yang2019,Todisco2020}. The index of SiO$_2$ is utilized as the substrate and assumed with a real constant refractive index $n_{\text{SiO}_2}$=1.45. For WS$_2$, a Lorentz oscillator model with the thickness of 0.618 nm is adopted from the experimental parameters by Li et al\cite{Li2014,Zhang_2018}:
\begin{equation}
	\varepsilon{(\omega)}=	\varepsilon_{\infty}+\frac{f\omega_{ex}^2}{\omega_{ex}^2-\omega^2-i\Gamma_{ex}^2\omega}.\label{eq15}
\end{equation}
where $\varepsilon_{\infty}=16$ is the background dielectric constant, $f=0.457$ represents the reduced oscillator strength, $\hbar\omega_0=2.013$ eV and $\hbar\Gamma_{ex}=22$ meV are the exciton resonance energies
and the exciton full line width, respectively. As depicted in Fig. \ref{fig1}(b), the most interesting finding is that the imaginary part appears a sharp peak (red line) around the transition energy of the exciton 2.01 eV. Therefore, WS$_2$ is a favourable option to realize the strong light-matter coupling.

The full-wave numerical simulations are performed with the finite element method via commercially available software COMSOL Multiphysics. In the following simulation calculations, the $y$-polarized plane wave is normally incident along the $z$ direction, where the periodic boundary conditions are adopted in both the $x$ and $y$ directions, and the perfectly matched layers are employed in the $z$ direction. 
  
\section{\label{sec3}Results and discussions}

In theory, symmetry-protected BIC with an infinite  $Q$ factor cannot be excited by a normally incident plane wave with any linear polarization\cite{Hsu2016,He2018} (see Figs. S1(a) and S1(b) in the Supplemental Material\cite{SM2022}); while for practical use, one can transform the BIC to a quasi-BIC by breaking the in-plane inversion symmetry of the system, where both the lifetime and line width become finite (see Figs. S2(a) and S2(b) in the Supplemental Material\cite{SM2022}).  Based on this, we introduce an off-centered hole in the nanodisk to open a radiation channel for obtaining an energy exchange with the external modes.

\begin{figure}[htbp]
	\centering
	\includegraphics
	[scale=0.36]{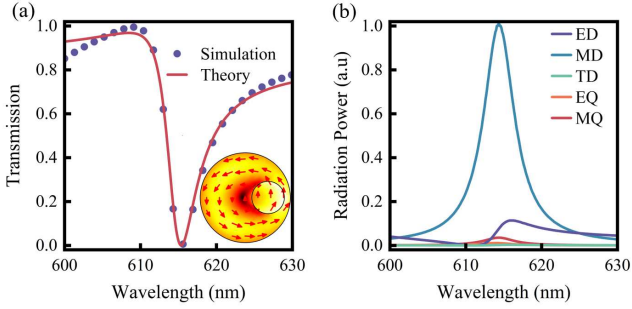}
	\caption{\label{fig2}(a) The simulated and theoretical transmission spectra of nanodisk metasurfaces and the inset shows the corresponding electric field distribution of $|E_{xy}|$ at the resonance, overlaid with arrows indicating the direction of the displacement current. (b) The scattered power of the multipole moments under the Cartesian cordinates, including the electric dipole (ED), magnetic dipole (MD), toroidal dipole (TD), electric quadrupole (EQ), and the magnetic quadrupole (MQ).}
\end{figure}

To engineer the radiation loss in the proposed nanodisk metasurfaces, we first analyze the resonant mode governed by the BIC mode without the presence of WS$_2$. Fig. \ref{fig2} (a) presents the simulated transmission spectrum (blue line) at a radius of the off-centered hole $r=50$ nm. we can find that it shows an asymmetric line-shape Fano resonance with a dip around 616 nm, and the simulation result is well fitted (red line) by the Fano formula\cite{Fan2003,Li2019,Xiao2020,Zhou2020},
\begin{equation}
 T(\omega)={\left| a_1+ja_2 + \frac{b}{\omega  - \omega _0 + j\gamma } \right|^2},\label{eq16}
\end{equation}
where $a_1$, $a_2$, $ b $ are the constant real numbers, $\omega_0$ is the resonant frequency, and $\gamma$ is the radiation loss of the quasi-BIC, which suggests that the quasi-BIC obtains an energy exchange with the continuum free-space radiation modes. The near field distribution is plotted in the inset of Fig. \ref{fig2} (a), where the bare metasurfaces hold stronger excitation of electric field at the center of the off-centered hole, and the distinct circular displacement current in the $x$-$y$ plane of the nanodisk suggests that the energy is strongly localized by the magnetic dipole moment oscillation along the $z$ direction\cite{Zhang2013}. In order to further identify the magnetic dipole quasi-BIC resonance, the electromagnetic multipoles are decomposed under the Cartesian coordinates in Fig. \ref{fig2}(b), and it is obvious that radiation contribution of the magnetic dipole ultimately dominates\cite{Kaelberer2010,Savinov2014,Terekhov2019}(More details can be found in the Supplemental Material\cite{SM2022}).

\begin{figure}[htbp]
	\centering
	\includegraphics
	[scale=0.36]{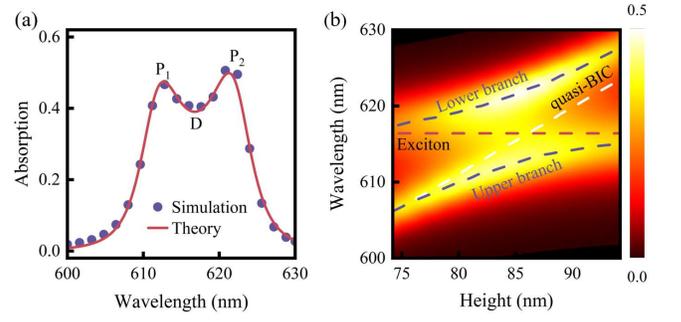}
	\caption{\label{fig3} (a) The simulated and theoretical absorption spectra of the coupled system with asymmetric $\alpha=0.127$ and $H=85$ nm. (b) The absorption spectra of the new hybrid state with different heights H and the asymmetric parameter $\alpha=0.127$. The red line and white line depict the individual exciton and quasi-BIC mode; blue dashed lines are the upper branch (UB) and lower branch (LB), respectively.}
\end{figure}

Next, we turn to the case where a monolayer WS$_2$ with a dissipate loss rate $\delta$ is spread on the nanodisk metasurfaces. As shown in Fig. \ref{fig3}(a), a clear spectral splitting with two peaks (P$_1$ and P$_2$ ) and one dip (D) is observed in the visible range from 600 nm to 630  nm. The simulated absorption curve can be well fitted (red line) by Eq. (\ref{eq14}) via the generalized TCMT. The coupling strength and Rabi splitting are found as 17.91 meV and 29.33 meV, respectively. The absorption loss $\delta=11$ meV was obtained from date by Li et al \cite{Li2014}, and the radiation rate $\gamma=7.44$ meV can be extracted from the Fig. \ref{fig2}(a). Significantly, ${g}>\frac{|\delta-\gamma|}{2}$ and ${g}>\frac{\sqrt {\delta^2+\gamma^2}}{2}$, which indicate that the strong coupling between the magnetic dipole quasi-BIC resonance and exciton is realized. It is noted that, this results dramatically differ from individual exciton or quasi-BIC mode, since the original energy levels are hybridized, and formed two new energy levels once the strong light-matter coupling condition is satisfied.

Fig. \ref{fig3}(b) presents the simulated optical behavior of the coupled system with different heights of the nanodisk. A typical anti-crossing behaviour can be captured in the 2D color diagram, which further confirms that strong coupling between the magnetic dipole quasi-BIC and exciton is reached, and significant upper branch (UB) and lower branch (LB) states in the hybrid system are finally formed. At zero detuning $\omega_\text{n}-\omega_\text{e}=0$, where the $\omega_\text{n}$ and the $\omega_\text{e}$ are resonance frequency of the quasi-BIC resonator and exciton, respectively, the Rabi splitting 29.33 meV can be extracted by the numerical simulation results shown in Fig. \ref{fig3}(b), which is matched well with the theoretical values predicted by our theoretical frame work.
\begin{figure}[htbp]
	\centering
	\includegraphics
	[scale=0.36]{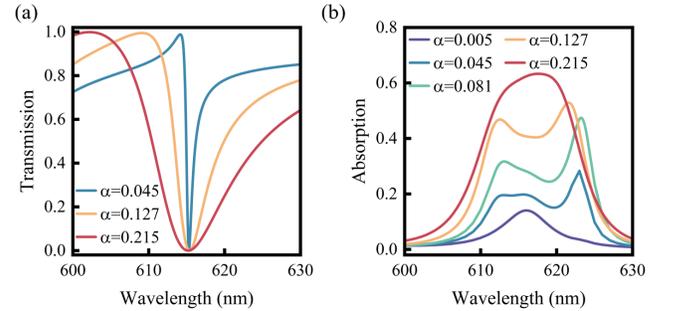}
	\caption{\label{fig4} (a) The simulated transimission spectra with different asymmetric parameters $\alpha$. (b) The evolution of the absorption line shape with the asymmetric parameter under the energy detuning $\omega_\text{n}-\omega_\text{e}=0$.}
\end{figure}

 In our work, the WS$_2$ is laid on the TiO$_2$ nanodisk metasurfaces. Hence, the light-matter coupling strength $g$ is proportional to the electric field at the surface of the metasurfaces. However, the higher factor $Q$ usually is accompanied by the lower radiation energy. As shown in Figs. \ref{fig4}(a) and 4(b), the simulated transimission and absorption curves  for different asymmetric parameter $\alpha $, resonant wavelength  all around 616 nm by tuning the thickness. When the asymmetric parameter  $\alpha=0.005$, $Q$=4.6$\times10$$^4$, the spectral line width is very narrow (only 0.96 meV), which is far less than the exciton line width, preventing the strong coupling phenomenon (the Rabi splitting is invisible). In contrast, for $\alpha=0.127$, $Q$=133, the quasi-BIC possesses a 14.88 meV line width, approaching exciton line width (22 meV), resulting a remarkable spectral splitting. Finally, when the asymmetric parameter $\alpha$=0.215, $Q$=52.2, the line width of quasi-BIC is about 38.56 meV, which is more than exciton line width and the eletric filed enhancement is very small at the interface, as a result of the absorption curve becomes a Lorentz-line shape. The reason for this unsuspected behavior is the simultaneous affection by the spectral line width and electric filed enhancement. Despite the local electric field is stronger for smaller $\alpha$, the spectral line width of the quasi-BIC resonance mode is more narrower, when it is far less than the exciton line width of 22 meV, the number of photons that can participate in the coupling effect is much smaller which degrades the Rabi splitting. Hence, Rabi splitting significantly increases with the number of photons that can participate in the coupling, and the balance between the line width and the local electric field enhancement of quasi-BIC mode needs to be carefully considered when designing the strong coupling system.

To further explore the physical mechanism behind the above phenomenon, we deduce the Eq. (\ref{eq14}) under the detuning condition, $\omega_\text{n}-\omega_\text{e}=0$ and the absorption of the hybrid system can be further expressed as
\begin{equation}
	A=\frac{2\delta \gamma {g}^2 }{[\delta \gamma+{g}^2]^2}.\label{eq15}
\end{equation}
It can be seen from the formula that there are only two factors that affect the absorption spectrum, the radiation loss and coupling strength, since the absorption loss of exciton is constant in the design. It is known that the coupling strength $g$ is proportional to the electric field intensity of the interface between the two-dimensional materials and the metasurfaces \cite{Chen2020}.  As mentioned earlier, the magnetic dipole moment oscillation along the $z$ direction, and the material is tiled on the surface of the metasurfaces in our proposed model, so the electric field strength of the contact surface should be related to the local field strength and radiation loss of the quasi-BIC resonance. It suggests that the strong coupling system can be manipulated by simply controlling the radiation loss. For example, when radiation loss is very small, Rabi splitting will gradually disappears and the coupled system enters the weak coupling region, in which the Purcell effect can be enhanced for the study of emitters\cite{Hugall2018}. As the radiation loss increases to supply enough photons, and the spectrum splits into two peaks and achieves the strong coupling, which can be used for quantum electrodynamics, nonlinear enhancement, and so on \cite{Kravtsov2020,Han2018}. However, when the radiation loss satisfies the critical condition $\text{g}^2=\delta\gamma$, the split spectrum evolves into Lorentz lines which can be used to enhance the absorption of 2D materials\cite{Piper2014,Xiao2020,Wang2020,Nie2020}.

\section{\label{sec4}Conclusions}

In conclusion, we have generalized the TCMT to study the strong light-matter coupling, in which all involved parameters can be directly computed by our fitting procedures. In an exemplary system composed of the monolayer WS$_2$ on the TiO$_2$ nanodisk metasurfaces, the strong coupling between the magnetic dipole quasi-BIC resonance and exciton is realized with the coupling strength up to 29.33 meV. Furthermore, it is important to balance the linewidth of the quasi-BIC mode and local electric field enhancement since both of them affect the strong coupling. Beyond this example, our  proposed analytical model has clear physical meaning, high fitting accuracy, and can be widely applied to various coupled systems composed of different resonant nanostructures and exciton in TMDCs throughout the spectrum, which is potential for future compact quantum information devices.

\begin{acknowledgments}	
This work is supported by the National Natural Science Foundation of China (Grants No. 11947065, No. 12064025), the Natural Science Foundation of Jiangxi Province (Grants No. 20202BAB211007, No. 20212ACB202006), the Interdisciplinary Innovation Fund of Nanchang University (Grant No. 20199166-27060003), the Open Project of Shandong Provincial Key Laboratory of Optics and Photonic Devices (Grant No. K202102), and the Major Discipline Academic and Technical Leaders Training Program of Jiangxi Province (Grant No. 20204BCJ22012).

M.Q. and J.D. contributed equally to this work.
\end{acknowledgments}


%

\end{document}